\documentclass[letter]{aa} 
\usepackage[switch]{lineno}
\usepackage[normalem]{ulem}
\usepackage{textcomp}
\usepackage{float}
\usepackage{dcolumn}
\newcolumntype{d}[1]{D{.}{.}{#1}}  
\usepackage{graphicx}
\usepackage{xcolor,colortbl}
\usepackage{booktabs}
\usepackage{txfonts}
\usepackage{natbib}
\usepackage{subfigure}
\usepackage{xcolor}
\usepackage{amsmath}
\usepackage[colorlinks=true,urlcolor=blue,citecolor=blue,linkcolor=blue,breaklinks=true]{hyperref}

\begin{document} 

\title{\ion{H}{i} envelope around the carbon star V420\,Vul}
\titlerunning{\ion{H}{i} envelope around V420\,Vul}
\authorrunning{Ouyang et al.}

\author{Xu-Jia Ouyang\inst{1} \thanks{xjouyang@gzu.edu.cn}
          \and
          Yong Zhang
          \inst{2,3}  \thanks{zhangyong5@email.sysu.edu.cn}
          \and
          Chuan-Peng Zhang\inst{4}
         \and
          Li-Yun Zhang\inst{1}
          }

\institute{College of Physics, Guizhou University, Huaxi District, Guiyang, Guizhou province, China  \email{xjouyang@gzu.edu.cn}
      \and
School of Physics and Astronomy, Sun Yat-sen University, 2 Daxue Road, Tangjia, Zhuhai 519082, Guangdong Province, China \email{zhangyong5@email.sysu.edu.cn}
      \and
          CSST Science Center for the Guangdong-Hongkong-Macau Greater Bay Area, Sun Yat-Sen University, Guangdong Province, China
        \and 
        State Key Laboratory of Radio Astronomy and Technology, National Astronomical Observatories, Chinese Academy of Sciences, Beijing 100101, China
          }

   \date{\today}

\abstract{
We report the detection of an extended 21-cm parsec-scale \ion{H}{i} structure toward the Mira variable V420\,Vul using archival Galactic Arecibo L-band Feed Array survey data. The emission exhibits a spatially coherent but intensity-asymmetric morphology that nevertheless retains a globally symmetric kinematic profile centered near $v_{\mathrm{LSR}} \sim 47.6\,\mathrm{km\,s^{-1}}$.
At an adopted distance of $\sim 1.9$\,kpc, the structure extends over $\sim 20$\,pc, implying a dynamical timescale on the order of $10^6$\,yr. 
Although the total \ion{H}{i} mass ($\sim 70\,\mathrm{M_\sun}$) indicates that the atomic gas reservoir is heavily dominated by the ambient interstellar medium rather than pristine stellar ejecta, the spatially resolved spectra and position-velocity diagrams reveal an underlying symmetric velocity framework centered on the star. We interpret this as the dynamical imprint of the stellar wind preferentially channeling through a porous ambient cloud, demonstrating that cohesive kinematic records of late-stage stellar mass loss can be preserved over parsec scales and megayear timescales despite dominant interstellar coupling.
}


\maketitle
\nolinenumbers

\section{Introduction} \label{intro}

During the asymptotic giant branch (AGB) phase, low- to intermediate-mass stars ($0.8-8\,\mathrm{M_\sun}$) undergo intense mass loss, returning large quantities of gas and dust to the interstellar medium (ISM). Characterizing the long-term evolution of this mass loss is crucial for understanding both the final stages of stellar evolution and the role of evolved stars in shaping the surrounding ISM.

A complete observational reconstruction of stellar mass-loss history remains a major challenge. The standard tracers of mass-loss history rely on observations of CO millimeter-wave line emission  or dust-scattered light. 
However, due to photodissociation by the interstellar ultraviolet (UV) radiation field, CO line emission can only trace out to a radius of $\sim 2\times10^{17}$\,cm ($\sim 0.06$\,pc) or less, even for stars with mass-loss rates as high as $10^{-5}$\,$\mathrm{M_\sun}$\,$\mathrm{yr}^{-1}$ \citep[e.g.,][]{2019A&A...625A..81S,2020A&A...640A.133R}. This corresponds to a mass-loss timescale of approximately 6500~years for a typical expansion velocity of $10$\,$\mathrm{km\,s^{-1}}$.
In contrast, circumstellar envelopes (CSEs) of AGB stars almost ubiquitously exhibit extended dust emission well traced by far-infrared (FIR) imaging, which has long been instrumental in characterizing historical, large-scale mass loss \citep[e.g.][]{1993ApJS...86..517Y, 2012A&A...537A..35C}. Even more extended, gas-dominated structures are accessible via far-ultraviolet (FUV) imaging from the Galaxy Evolution Explorer (GALEX), tracing extended spatial scales. Most notably, it reveals a trail spanning $2^\circ$ ($\sim 4$\,pc) around $o$\,Ceti \citep{2007Natur.448..780M} and across numerous other evolved stars \citep[e.g.][]{2023AJ....165..229S,2023A&A...680A..12R}. 
However, despite their large spatial extent, both FIR continuum and FUV imaging inherently lack kinematic information, making it impossible to directly constrain the expansion velocities and spatial dynamics of these outer envelopes.

Atomic hydrogen (\ion{H}{i}) constitutes a fundamental component of evolved stellar outflows \citep{1983MNRAS.203..517G} and serves as a key tracer of stellar mass loss. However, reconstructing mass-loss histories via \ion{H}{i} emission remains challenging, owing to strong contamination from interstellar 21-cm line emission and the intrinsically faint circumstellar \ion{H}{i} signal.
The deployment of the Nançay Radio Telescope (NRT) and the Very Large Array (VLA) to the study of circumstellar \ion{H}{i} has driven significant progress in this field \citep[e.g.,][]{2010A&A...515A.112L,2012MNRAS.422.3433L,2024A&A...692A..54G,2006AJ....132.2566G,2008ApJ...684..603M,2013AJ....145...97M,2014A&A...565A..54H}.
Nevertheless, single-dish observations with the NRT are heavily limited by beam confusion from Galactic interstellar \ion{H}{i} emission due to its highly elongated beam. Conversely, while interferometers such as VLA provide high-resolution imaging, they inherently suffer from a lack of sensitivity to extended, low-surface-brightness circumstellar structures.
Bound by these instrumental limitations, past \ion{H}{i} surveys typically achieved detections limited to the densest, inner CSEs, with physical scales rarely exceeding a few parsecs \citep[e.g.,][]{2006AJ....132.2566G,2024A&A...692A..54G}, thereby hindering comprehensive reconstructions of long-term mass-loss histories. Overcoming these biases requires large-aperture single-dish telescopes capable of balancing high sensitivity with sufficient angular resolution. Although Green Bank Telescope did indeed demonstrate the value of mapping extended outflows \citep{2015MNRAS.449..220M}, its $\sim 9\arcmin$ beam ends up blurring the picture of the detailed kinematics involved.
Crucially, while archival data from the Arecibo Observatory—specifically Galactic Arecibo L-band Feed Array \ion{H}{i} (GALFA–\ion{H}{i}) survey \citep{2018ApJS..234....2P} utilized in the present work—provide a valuable legacy for exploring extended envelopes, Five-hundred-meter Aperture Spherical Telescope (FAST) now offers unprecedented potential to systematically map such diffuse environments.

In this Letter, we present the detection of a giant, $\sim 40\arcmin$ asymmetric \ion{H}{i} envelope around the Mira variable V420\,Vul, using archival data from the GALFA-\ion{H}{i} survey. Details of the data and reduction are provided in Appendix \ref{sec:obse}.

V420\,Vul ($l=71.88^\circ$, $b=-12.67^\circ$, epoch J2000) is a hitherto poorly studied carbon-rich (C-type) Mira variable, with a well-measured pulsation period of 377~days \citep{2017ARep...61...80S}.
Prior to this work, the most detailed study of this source was presented by \citet{2023A&A...680A..12R}, who detected faint extended FUV emission spanning $5\arcmin$--$6\arcmin$ around the star using archival GALEX data. From radiative transfer modelling with the \texttt{DUSTY} code, the authors derived the star’s current-day fundamental parameters: an effective temperature of 3000\,K, a luminosity of $4000\,\mathrm{L}_\sun$, and a present-day mass-loss rate of $1.7\times10^{-7}\,\mathrm{M_\sun}\,\mathrm{yr}^{-1}$.

\section{Results} \label{sec:resu}

Figure~\ref{fig:cm} (provided in Appendix~\ref{cm}) displays channel maps covering the velocity range $30$--$70\,\mathrm{km\,s^{-1}}$ \footnote{ Unless otherwise noted, all velocities quoted throughout this paper are in the Local Standard of Rest (LSR) frame.}. Clear \ion{H}{i} emission spatially overlapping with the position of V420\,Vul is detected over the velocity range $36$--$68\,\mathrm{km\,s^{-1}}$. 
As discussed in Appendix~\ref{dis}, the systemic velocity of V420\,Vul remains uncertain due to poor astrometry and strong atmospheric pulsations of long-period variables (LPVs). The channel maps reveal a large-scale spatial elongation along the northwest--southeast (NW--SE) axis across the entire $36$--$68\,\mathrm{km\,s^{-1}}$ range, indicating a spatially anisotropic but kinematically coherent and symmetric structure. We discuss its physical origin is further in Sect.~\ref{sec:discu}.

\begin{figure*}
\includegraphics[width=2.0\columnwidth]{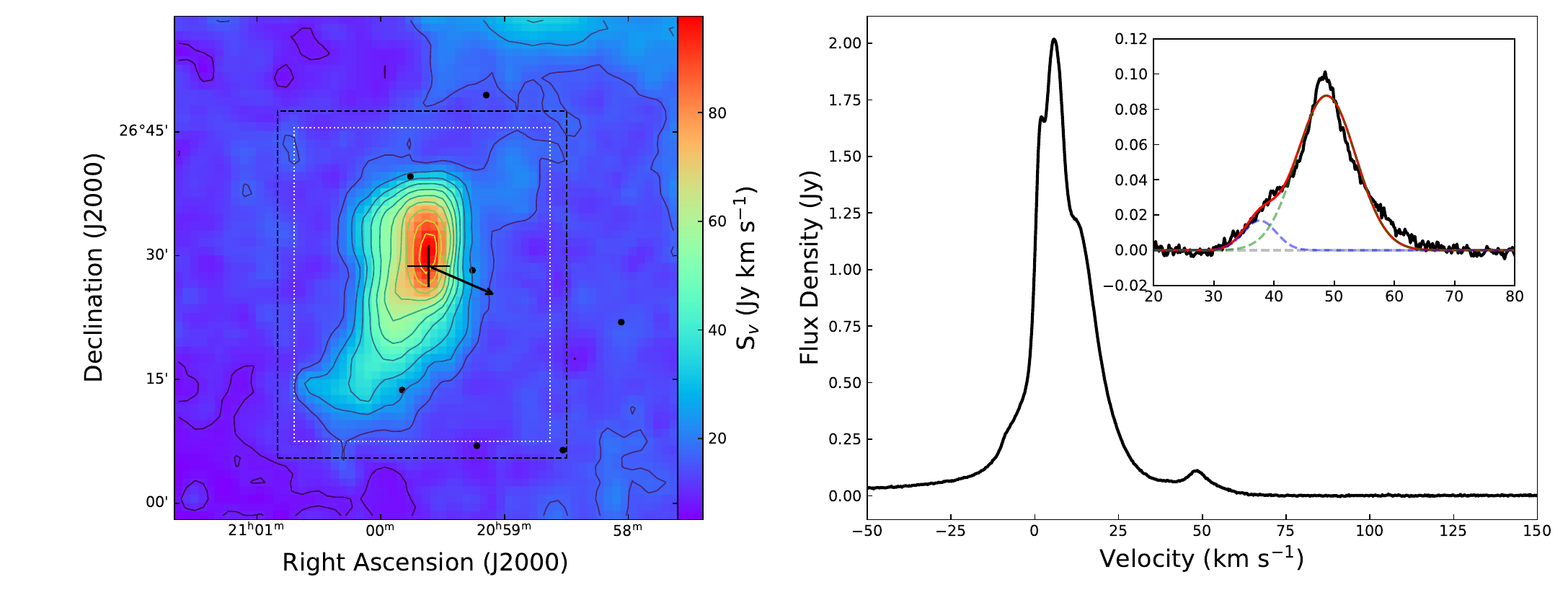}
\caption{
{\it Left}: Integrated \ion{H}{i} intensity map over $36$--$68\,\mathrm{km\,s^{-1}}$. Contours start at $3\sigma$ and increase in steps of $3\sigma$. The outermost stellar-associated contour begins at $6\sigma$. The plus sign marks V420\,Vul, with an arrow indicating its proper motion. The black dots denote SIMBAD LPVs within the region, plotted after the filtering described in Sect.~\ref{sec:discu}. The white box encloses the region used for \ion{H}{i} mass calculation, and the area between the black and white boxes defines the background region. {\it Right}: 
Averaged \ion{H}{i} spectrum extracted from the white box in the left panel. The upper right inset zooms in on the spectral feature across the $20$--$80\,\mathrm{km\,s^{-1}}$ velocity range with a two-component Gaussian fit overlaid. The solid red curve denotes the composite fit. The solid orange curve and green dashed curve represent the primary ($\sim49\,\mathrm{km\,s^{-1}}$) and secondary ($\sim39\,\mathrm{km\,s^{-1}}$) Gaussian components, respectively, while the gray dashed horizontal line marks the zero baseline.
}
\label{fig:m0}
\end{figure*}

We integrated the \ion{H}{i} emission over $36$--$68\,\mathrm{km\,s^{-1}}$ to construct the integrated intensity map (Fig.~\ref{fig:m0}, left panel), yielding a root-mean-square (rms) noise of $\sim 2.8\,\mathrm{Jy\,km\,s^{-1}}$ evaluated across emission-free regions.
We then extracted the spatially averaged spectrum (right panel of Figure~\ref{fig:m0}) by taking the mean of the spectral pixels within the white rectangular region marked in the intensity map. After the baseline subtraction, the integrated spectrum exhibits an asymmetric profile with a mild double-peaked appearance. This likely reflects spatial variations in velocity across the emitting region rather than a single homogeneous kinematic structure. This behavior and its spatial variation are discussed in Appendix~\ref{sec:grid}.
We derived a background-subtracted integrated \ion{H}{i} flux of $\int S_\nu \mathrm{d}v = 82.7 \pm 1.8\,\mathrm{Jy\,km\,s^{-1}}$ from the white rectangular region shown in Figure~\ref{fig:m0}, where the background level is estimated from the region enclosed between the white and black rectangles.

\begin{figure}
\includegraphics[width=1.0\columnwidth]{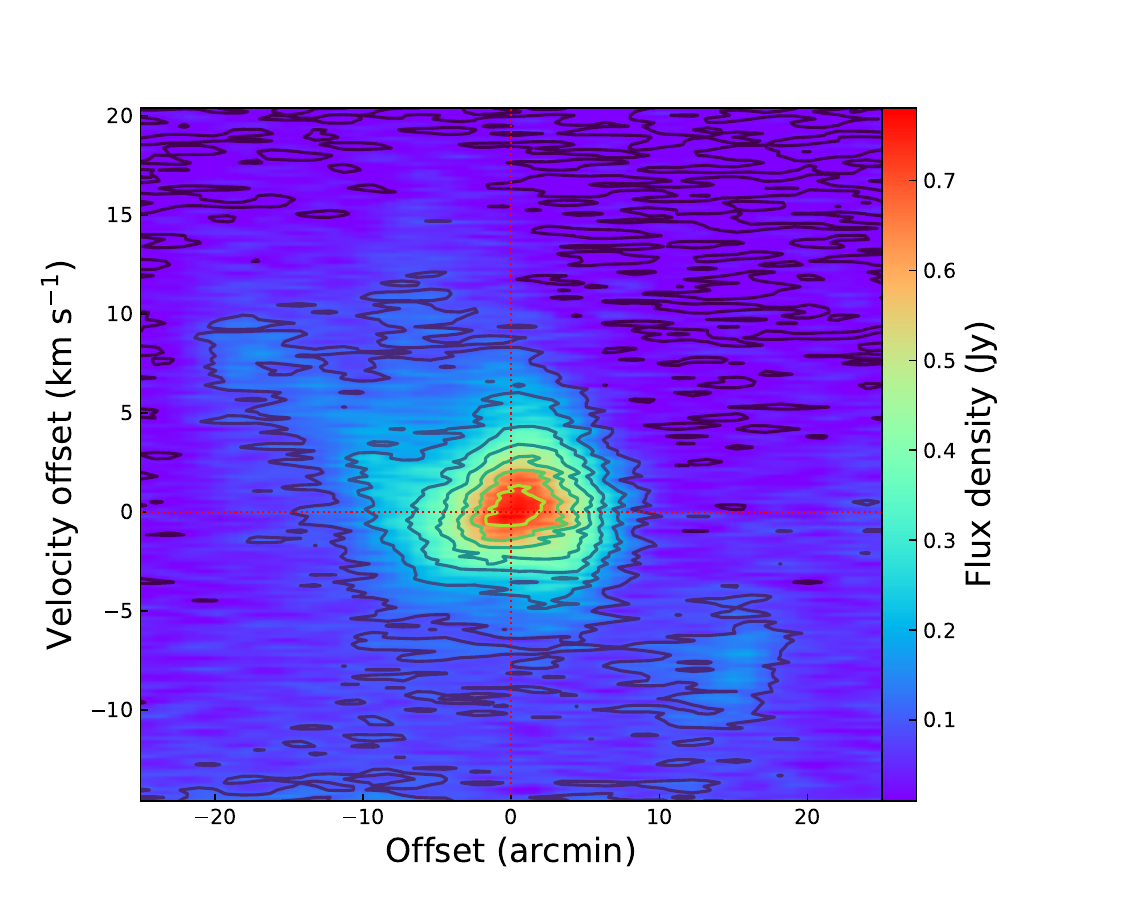}
\caption{PV diagram of \ion{H}{i} emission along position angle
$\rm PA=145^{\degr}$. The origin $(0, 0)$ corresponds to the position of V420\,Vul and an LSR reference velocity of $V_{\rm LSR} = 47.6\,\mathrm{km\,s^{-1}}$ (indicated by the dashed red lines).
}
\label{fig:pv}
\end{figure}

A position–velocity (PV) diagram (Figure~\ref{fig:pv}) was extracted along the major axis of the emission, adopting a position angle of $145^\circ$.  We took a representative systemic velocity of $v_{\mathrm{LSR}} \sim 47.6\,\mathrm{km\,s^{-1}}$ based on the centroid of the integrated profile and the approximate symmetry of the outer envelope, while noting potential contamination from ambient interstellar emission.
Crucially, the PV structure reveals a faint but kinematically coherent extension reaching velocity offsets of up to $\sim 8\,\mathrm{km\,s^{-1}}$ relative to the systemic velocity, comparable to typical AGB wind expansion speeds \citep{2018A&ARv..26....1H}. This low-level emission is superimposed on a brighter central concentration that is largely confined within $\sim 5\,\mathrm{km\,s^{-1}}$ of the systemic velocity.

Under the optically thin assumption and adopting the period-luminosity relation (PLR) distance of $\sim 1.9$\,kpc, we derived a total background-subtracted \ion{H}{i} mass of $\sim 70\,\mathrm{M_\sun}$, which represents the nominal net mass after mitigating line-of-sight Galactic blending. 
The characteristic spatial scale of $\sim 20$\,pc implies a dynamical timescale on the order of $10^6$\,yr, assuming that the observed \ion{H}{i} structure is physically connected and dynamically driven by the central star.
This structure is significantly more extended than comparable \ion{H}{i} envelopes, such as the $\sim0.8$\,pc structure around IRC+10216 \citep{2015MNRAS.449..220M}, and the typical parsec-scale structures detected around evolved stars in previous studies, which characteristically average $\sim 1.6$\,pc and rarely exceed $\sim 5$\,pc \citep[e.g.,][]{2024A&A...692A..54G}. Despite the large spatial scale, the inferred timescale remains broadly consistent with late-stage AGB evolutionary timescales within an order of magnitude \citep{2018A&ARv..26....1H}.

\section{Discussion} \label{sec:discu}

\ion{H}{i} emission detected in CSEs can arise from three distinct mechanisms: a predominantly atomic stellar wind (for $T_{\rm eff}\gtrsim2500$\,K; \citealt{1983MNRAS.203..517G}), swept-up interstellar gas, or the destruction of $\mathrm{H}_2$ via shock-driven or radiative photo-dissociation. 
The observed \ion{H}{i} morphology in V420\,Vul disfavors a classical swept-up ISM shell. Although faint extended FUV emission is present in the \textit{GALEX} image \citep{2023A&A...680A..12R}, it shows no discernible bow-shape feature and provides no evidence for the high-velocity shocks or strong UV radiation fields required to produce any substantial $\mathrm{H}_2$ dissociation.
Given a value of $T_{\rm eff}\approx3000$\,K \citep{2023A&A...680A..12R}, the stellar wind is expected to be intrinsically atomic. Nevertheless, the exceptionally large inferred \ion{H}{i} mass ($\sim70\,\mathrm{M_\sun}$) far exceeds expectations for a purely CSE, indicating that the atomic gas reservoir is heavily dominated by the ISM. Consequently, we interpret this system as an embedded evolutionary framework, where the star resides within a pre-existing diffuse interstellar cloud; within this framework, the ambient cloud provides the primary mass reservoir, whereas the stellar wind acts as the primary dynamical driver that continuously shapes the surrounding gas and imprints the distinctive large-scale velocity spread observed across the medium. While a line-of-sight coincidence cannot be fully excluded, the spatial–kinematic coherence across multiple tracers makes such a scenario unlikely.

The PV diagram shows a broad spatial–kinematic coherence centered on the position of V420\,Vul, suggesting a physical connection between the stellar source and part of the surrounding \ion{H}{i} structure. To further secure this association and rule out potential line-of-sight coincidences common in complex Galactic \ion{H}{i} fields, we examined archival IRAS and AKARI imaging ($25$--$140\,\mu$m; Figure~\ref{fig:ir}). At $25\,\mu$m, V420\,Vul appears as an isolated compact source, with no nearby infrared-bright objects coincident with the \ion{H}{i} peak. A SIMBAD cross-match (excluding non-mass-losing sources) also reveals no nearby active LPVs spatially coincident with the \ion{H}{i} emission peak (see black dots in the left panel of Figure~\ref{fig:m0}). The absence of alternative mass-losing stellar sources strongly supports a physical connection between V420\,Vul and the observed 21-cm structure \citep[e.g.,][]{1988ApJ...328..763L,2016ARA&A..54..491B}, although minor contributions from unrelated diffuse interstellar material along the line of sight cannot be entirely eliminated.

The FIR ($60$--$140\,\mu$m) emission displays a north–south morphology broadly aligned with the \ion{H}{i} structure, while suggesting an apparent brightness anti-correlation: the northern FIR region is significantly enhanced, while the co-spatial \ion{H}{i} emission is noticeably suppressed. 
Although the limited spatial resolutions of both the FIR and \ion{H}{i} data prevent us from establishing whether the apparent positional offset between their emission peaks is statistically significant, this behavior is consistent with dust–gas coupling within a cold, inhomogeneous ambient medium.
In particular, an enhanced dust column density can promote efficient $\mathrm{H}_2$ formation on grain surfaces, thereby reducing the local atomic hydrogen abundance \citep{1983MNRAS.203..517G,2017MolAs...9....1W}. Concurrently, the substantial dust mass and low temperature derived from AKARI photometry ($M_{\rm d}\sim0.8\,\mathrm{M_\sun}$, $T_{\rm d}\approx19.4$\,K) might contribute to \ion{H}{i} self-absorption (HISA) under suitable conditions. The combined effect of these processes could suppress the observed 21-cm brightness toward dense dust clumps, allowing the peak \ion{H}{i} emission to arise from relatively low-opacity regions between FIR enhancements. Within this framework, the observed $\sim1.8\arcmin$ ($\sim1$\,pc at 1.9\,kpc) offset between the star and the \ion{H}{i} intensity peak may result from the combined influence of dynamical evolution and radiative transfer effects: while wind--ISM interactions or ram-pressure stripping shape the underlying gas distribution, localized $\mathrm{H}_2$ conversion, and possible HISA within the inhomogeneous ambient medium could further modulate the observed 21-cm intensity distribution, potentially shifting the apparent brightness peak away from the stellar position.

The global \ion{H}{i} morphology is elongated along the NW--SE axis, exhibiting a large-scale spatial asymmetry where the emission extends significantly further to the south than to the north, even though the velocity structure in the PV diagram shows a remarkably symmetric velocity spread about the stellar position and systemic velocity. Over parsec scales and megayear timescales, such a configuration can persist if the outflow preferentially propagates through low-density channels in a porous medium without globally accelerating the cloud—an effect that is qualitatively analogous to flow channeling in simulations of jets interacting with a multiphase ISM \citep{2012ApJ...757..136W}. 
A coherently rotating disk is disfavored given its exceptionally large spatial scale. Instead, the velocity structure is more consistent with an embedded stellar wind interacting with a diffuse interstellar cloud. 
Specifically, while the high-intensity core in the P-V diagram remains localized near the spatial and velocity zero-points, the low-intensity peripheral features reveal an extended, quasi-bipolar kinematic framework that is remarkably symmetric about the stellar position, propagating through low-density channels to extend coherently toward both the redshifted and blueshifted regimes. This symmetric velocity distribution directly reflects the dynamical footprint of stellar feedback expanding into a structured ambient cloud.
This scenario is further supported by the channel maps shown in Figure~\ref{fig:cm}. The southeastern emission spans a broad velocity range and is detected on both sides of the adopted systemic velocity ($v_{\rm LSR} \sim 47.6\,\mathrm{km\,s^{-1}}$). This components extends from the blueshifted regime well into the redshifted velocities, providing clear evidence for a blending between stellar wind motions and ambient cloud kinematics. Independently, the northwestern counterpart ($38$--$46\,\mathrm{km\,s^{-1}}$, position angle $\sim -50^\circ$) reveals subtle structural and kinematic distortions at lower emission levels, where the expanding wind directly encounters the highly inhomogeneous and porous ISM while the bulk velocity of the main cloud remains coherent.
As a consistency check for this embedded environment, adopting a canonical gas-to-dust ratio of $\sim100$ \citep{1978ApJ...224..132B} yields a dust-inferred gas mass of $\sim80\,\mathrm{M_\sun}$. Given this result is broadly consistent within the large uncertainties associated with the \ion{H}{i}-derived value ($\sim70\,\mathrm{M_\sun}$), it suggests that the diffuse ambient cloud contributes the majority of the total mass budget.

The observed velocity amplitudes of $\sim8\,\mathrm{km\,s^{-1}}$ (uncorrected for projection) are comparable to typical AGB wind speeds, but significantly lower than the high-velocity regimes ($50\text{--}100\,\mathrm{km\,s^{-1}}$) characteristic of fast post-AGB jets \citep{2025IAUS..384..337S}, suggesting that the system has not undergone strong post-AGB acceleration. Even when considering potential deceleration as the stellar outflow couples with the ambient medium via low-density channels, the driving source remains hydrodynamically modest. 
Rather than a highly collimated or disrupted bipolar jet, the PV diagram reveals an extended, kinematically symmetric core-envelope structure centered on the spatial and velocity origins, characterized by a continuous and ordered spatial–velocity extension. While the elevated Gaia renormalised unit weight error (RUWE) value (1.8) could hint at unresolved astrometric complexity from a potential binary companion, alternative explanations, such as stellar variability effects, cannot be excluded \citep{2022MNRAS.513.2437P}.
Overall, V420\,Vul appears to be embedded in a large-scale interstellar structure whose kinematics are globally symmetric and directly influenced by the stellar wind, rather than representing a classical isolated CSE. Although a definitive decomposition of the pristine circumstellar component is beyond the scope of the present work, future high-resolution interferometric observations will be essential to disentangling the pristine stellar wind from the dominant ambient emission.

\section{Conclusion} \label{sec:concl}

We report the detection of an extended $\sim20$\,pc \ion{H}{i} structure toward the Mira variable V420\,Vul using archival GALFA-\ion{H}{i} data. The system exhibits a coherent large-scale kinematic structure with a characteristic velocity spread of $\sim 8\,\mathrm{km\,s^{-1}}$ over a dynamical timescale of $\sim 10^6$\,yr. Although the total \ion{H}{i} mass is dominated by the ambient ISM, the gas emission shows a spatially and kinematically coherent distribution centered on the position and systemic velocity of V420\,Vul. This suggests that stellar feedback can sustain coherent dynamical structures even when embedded in a structured ambient interstellar cloud.
The presence of a kinematically symmetric velocity field over parsec scales suggests that anisotropic density structures in the ambient medium—such as a porous or inhomogeneous environment—might allow for stellar winds to preferentially propagate through low-density channels. Coupled with local dust chemistry and HISA effects, these mechanisms collectively give rise to the observed large-scale intensity asymmetries while preserving the underlying kinematic symmetry over megayear timescales. No unique, isolated CSE morphology can be distinguished; instead, the observed \ion{H}{i} structure represents a coupled system in which stellar mass loss and interstellar material are dynamically intertwined. These results highlight the potential of wide-field \ion{H}{i} observations to probe long-term wind--ISM interactions and to preserve the kinematic signatures of evolved stars well beyond the classical circumstellar regime.

\begin{acknowledgements}

We thank the anonymous reviewer for insightful suggestions, which have significantly improved the quality of this work.
The financial supports of this work are from  the National SKA Program of China (Grant No. 2025SKA0120100),
the National Natural Science Foundation of China (NSFC, No. 12473027,  12333005, and 11973099),
the Guangdong Basic and Applied Basic Research Funding (No.\,2024A1515010798), the Guizhou University Natural Science Special Research Fund (Special Post, project No. 202612), the Basic Research Program Youth Guidance Project of Guizhou Province (Grant No. QN [2026] 001), and Guizhou Provincial Major Scientific and Technological Program XKBF (2025)011.
This work makes use of data from the Galactic ALFA \ion{H}{i} (GALFA-\ion{H}{i}) survey, obtained with the Arecibo 305-m Telescope and the Arecibo L-band Feed Array (ALFA). The GALFA-\ion{H}{i} survey was supported by the U.S. National Science Foundation.
\end{acknowledgements}

\bibliography{references}{}
\bibliographystyle{aa}

\begin{appendix}

\section{Data and reduction} \label{sec:obse}

The GALFA-\ion{H}{i} survey is a large-scale \ion{H}{i} mapping survey carried out with the L-Band Feed Array receiver on the Arecibo Observatory 305 m telescope \citep{2018ApJS..234....2P}.
The survey covers the full range of Right Ascension (RA) and a Declination (Dec) range of $-1^\circ20^\prime < \mathrm{Dec} < 38^\circ02^\prime$, with an angular resolution of $\sim 4\arcmin$.
The data are delivered in the position–position–velocity (PPV) domain.
We use the second data release (DR2) of the GALFA-\ion{H}{i} survey for this analysis.
The DR2 data are split into two sets of data cubes: the Wide and Narrow cubes, with velocity resolutions of $0.754\,\mathrm{km\,s^{-1}}$ and $0.184\,\mathrm{km\,s^{-1}}$, respectively. The cubes are resampled onto a RA-Dec Cartesian grid with $1\arcmin \times 1\arcmin$ pixels, where the physical pixel size varies with Declination due to the sky projection.

For this analysis, we exclusively utilize the Narrow cubes, as their superior velocity resolution ($\Delta v = 0.184\,\mathrm{km\,s^{-1}}$) fully resolves the characteristic expansion velocity of the circumstellar outflow around V420\,Vul. The dataset exhibits a standardized rms noise of $\sim 150\,\mathrm{mK}$ per $1\,\mathrm{km\,s^{-1}}$ velocity channel.

The GALFA-\ion{H}{i} archival data are calibrated in units of brightness temperature ($T_B$), which can be converted to flux density ($S_\nu$) via the Rayleigh–Jeans approximation, 
\begin{equation}
S_\nu = \frac{2 k_B \nu^2}{c^2} T_B \Omega,
\end{equation}
where $k_B$ is the Boltzmann constant, $\nu$ is the observed frequency, $c$ is the speed of light, and $\Omega$ is the solid angle of the telescope’s primary beam.
For the GALFA-\ion{H}{i} instrumental setup, this simplifies to the standard conversion relation:
\begin{equation}
S_\nu\ (\mathrm{Jy}) \approx 0.07 \times T_B\ (\mathrm{K}).
\end{equation}

\section{Channel maps} \label{cm}

\begin{figure*}
\includegraphics[width=2.0\columnwidth]{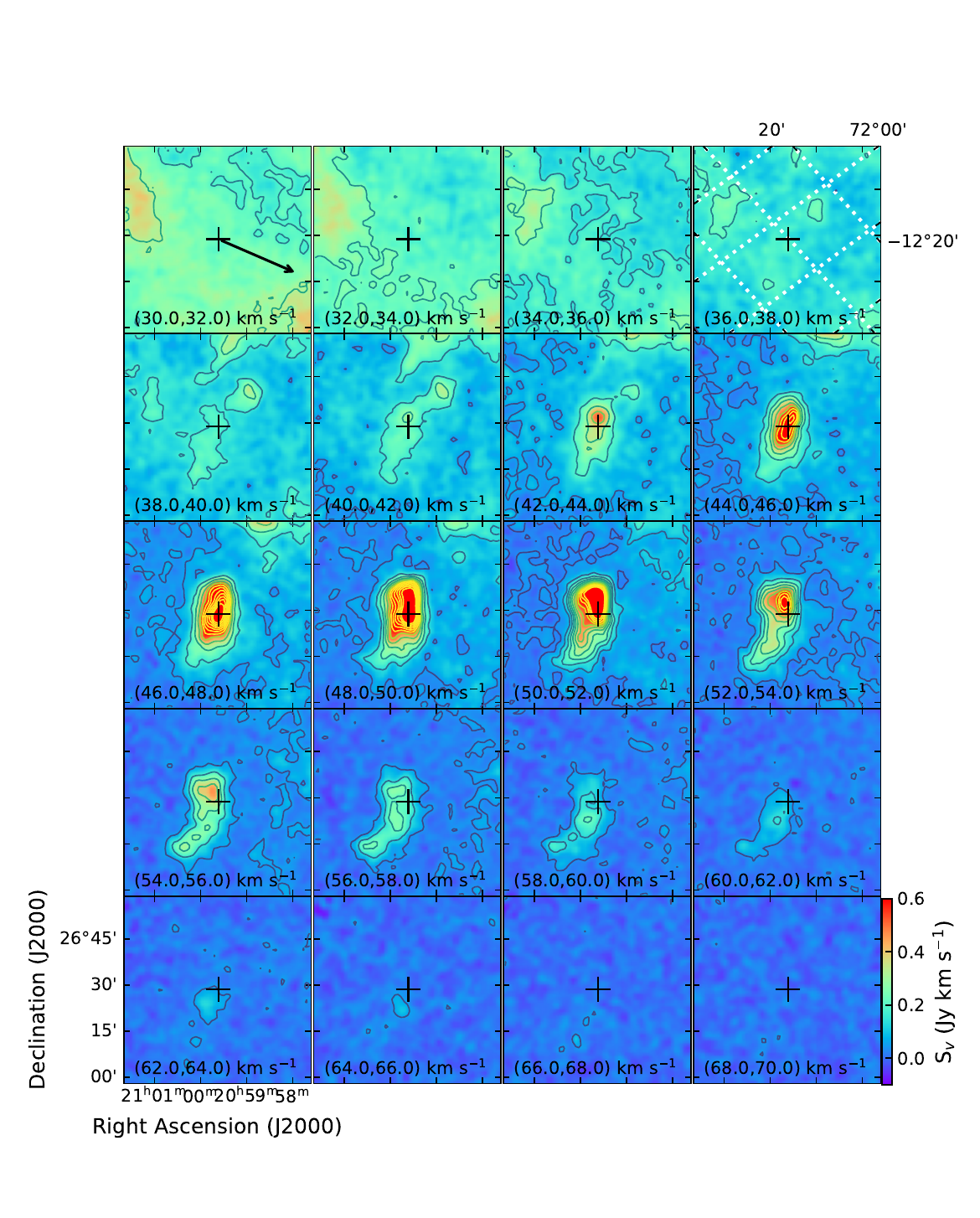}
\caption{
Channel maps of the \ion{H}{i} line in the LSR velocity range $v_{\mathrm{LSR}} = 30$ to $70\,\mathrm{km\,s^{-1}}$, with a channel width of $2\,\mathrm{km\,s^{-1}}$. The plus sign represents the position of V420\,Vul, and the arrow in the upper left panel indicates the direction of its proper motion. The galactic coordinate system is indicated by the white dotted grid lines in the upper right panel. Contour levels start at $3\sigma$ in each channel map, in steps of $6\sigma$.
}
\label{fig:cm}
\end{figure*}

Figure~\ref{fig:cm}: \ion{H}{i} channel maps of V420\,Vul. Each panel represents the integrated \ion{H}{i} intensity over the velocity interval indicated in the lower-left corner.

\section{Astrometric reliability} \label{dis}

According to the Gaia Data Release 3 (DR3) catalogue, the distance to V420\,Vul is $2.43^{+0.30}_{-0.24}$\,kpc \citep{2021A&A...649A...1G}. However, the reliability of this astrometric solution warrants caution. The source has a RUWE of 1.8, significantly exceeding the recommended threshold of 1.25 for well-behaved five-parameter astrometric solutions in Gaia DR3 \citep{2022MNRAS.513.2437P}. Such elevated RUWE values indicate that the standard single-source astrometric model does not adequately reproduce the observations and may arise from unresolved multiplicity, intrinsic photocenter variability, or other forms of astrometric complexity \citep{2018A&A...616A...2L}. Consequently, the formal parallax uncertainty may underestimate the true systematic uncertainty of the distance measurement.

Because the five astrometric parameters are solved simultaneously in the Gaia astrometric model, the same systematic effects responsible for the elevated RUWE may also affect the proper-motion solution. This is particularly relevant if the source hosts an unresolved long-period companion, for which orbital photocenter motion can bias both the measured parallax and the apparent proper motion over the Gaia observing baseline. Unlike the parallax, the induced proper-motion deviations can become substantial for orbital periods longer than about one year and may reach up to $\sim10$\,mas\,yr$^{-1}$ in extreme cases \citep{2020MNRAS.495..321P}. Although the binary nature of V420\,Vul remains unconfirmed, its elevated RUWE suggests that the published astrometric parameters, particularly the proper motion, should be interpreted with caution. We therefore refrain from deriving absolute transverse velocities from the Gaia proper motion. Instead, throughout this work the proper-motion vector is used only as a qualitative indicator of the projected direction of motion when comparing with the morphology of the \ion{H}{i} structure.

Nevertheless, no independent distance determination for V420\,Vul is currently available in the literature.
As an independent distance estimate less sensitive to astrometric systematics, we adopt the empirical bolometric PLR calibrated for C-type AGB stars \citep[Eq. 10;][]{2022A&A...667A..74A}: $M_{\mathrm{bol}} = -3.31 (\log P - 2.5) - 4.317$. With the well-established pulsation period of 377~days \citep{2017ARep...61...80S}, this yields an absolute bolometric magnitude $M_{\mathrm{bol}} \approx -4.57$. Combining this absolute magnitude with the observed near-infrared photometry ($J=5.666$, $K=3.793$; \citealt{2003yCat.2246....0C}) and appropriate bolometric corrections yields a PLR distance of $\sim 1.9\,\mathrm{kpc}$.

Despite the small formal uncertainty ($v_{\mathrm{helio}} = 14.86 \pm 1.55\,\mathrm{km\,s^{-1}}$) reported in Gaia DR3 \citep{2023A&A...674A...1G}, this heliocentric value corresponds to $v_{\mathrm{LSR}} \approx 30.6\,\mathrm{km\,s^{-1}}$, which lies outside the main channel map emission but slightly overlaps with the blue wing in the Fig.~\ref{fig:m0} (right) inset. Furthermore, this value may not accurately represent the stellar systemic velocity. According to the Gaia RVS pipeline design \citep{2023A&A...674A...5K}, accurate velocity measurements strictly depend on high-precision astrometry, as an error on the along-scan field angle propagates linearly to the wavelength zeropoint. Given that V420\,Vul exhibits a degraded astrometric solution with a high $\text{RUWE} = 1.8$, this linear error propagation can introduce non-negligible systematic shifts. Furthermore, for long-period variables, the observed radial velocities are heavily modulated by strong atmospheric pulsations. Consequently, a simple median derived from sparse-epoch observations \citep{2023A&A...674A...5K} might not fully capture the true systemic velocity. This may contribute to the observed $\sim17\,\mathrm{km\,s^{-1}}$ discrepancy relative to our gas-derived value.

\section{Grid analysis of the \ion{H}{i} spectral profiles} \label{sec:grid}

\begin{figure*}
\includegraphics[width=2.0\columnwidth]{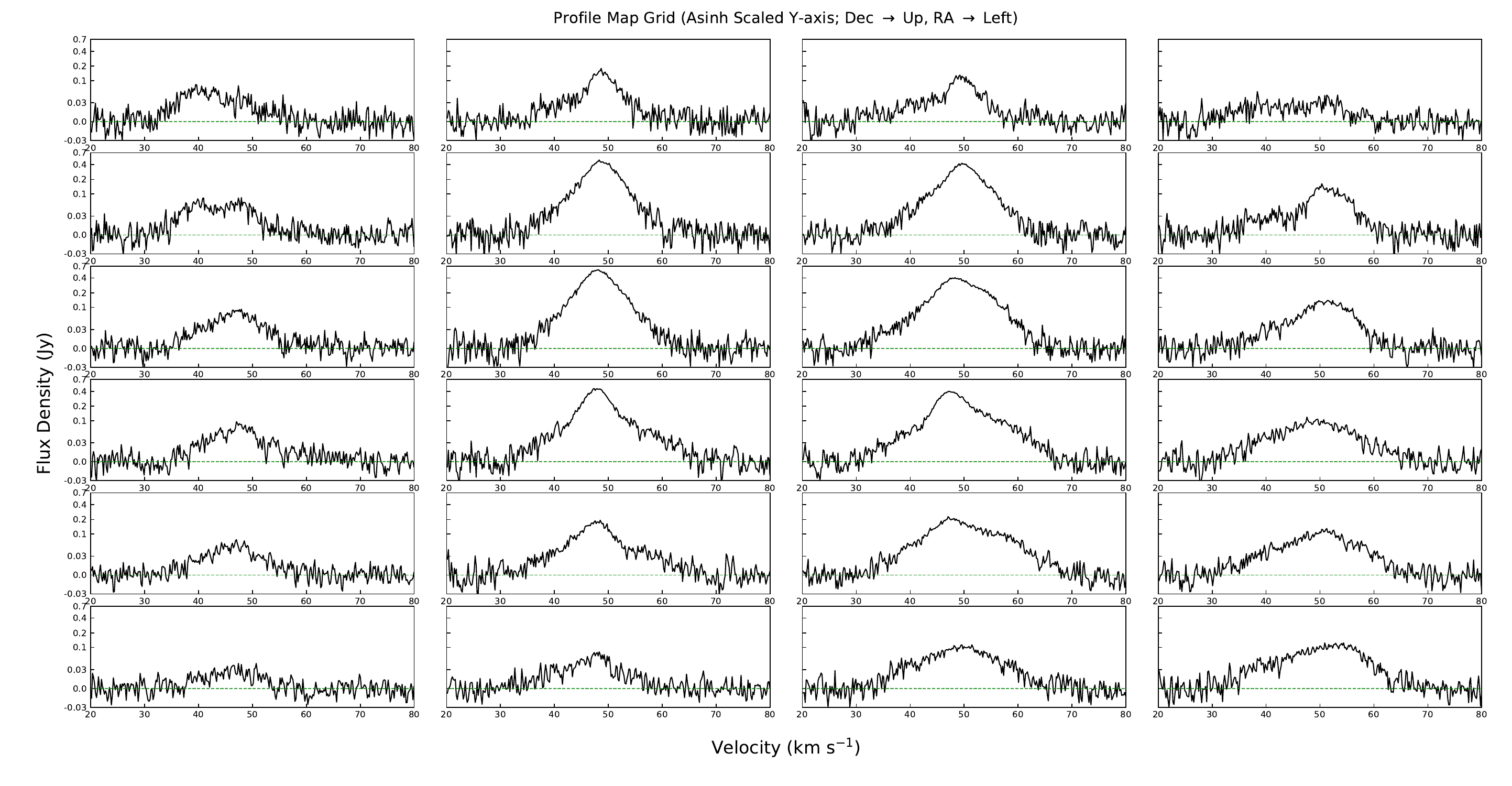}
\caption{Profile map grid of extracted spectra across a $20^{\prime} \times 30^{\prime}$ region centered on V420\,Vul. Each panel displays the spectrum extracted over an individual $5^{\prime} \times 5^{\prime}$ subregion. The grid layout corresponds to the spatial orientation on the sky, with Declination (Dec) increasing upward and Right Ascension (RA) increasing to the left. In each panel, the black solid line traces the flux density (in units of Jy) as a function of line-of-sight velocity (in $\mathrm{km\,s^{-1}}$), with the $y$-axis scaled using an inverse hyperbolic sine ($\text{asinh}$) function to emphasize faint features. The green dotted line denotes the zero-flux baseline.}
\label{fig:grid}
\end{figure*}

To examine whether the integrated \ion{H}{i} line profile arises from spatially distinct velocity components or from a coherent large-scale kinematic structure, we extracted spectra over a grid of $5^{\arcmin} \times 5^{\arcmin}$ regions across a $20^{\arcmin} \times 30^{\arcmin}$ area centered on V420\,Vul (see Fig.~\ref{fig:grid}).

Most individual spectra are characterized by a single broad emission component centered near $v_{\mathrm{LSR}} \approx 47.6\,\mathrm{km\,s^{-1}}$. While weak asymmetries or shoulders dominate the departures from single-Gaussian profiles, distinct double-peaked structures are limited to only a few isolated positions across the mapped region.

Instead, the peak velocity profiles exhibit a high degree of spatial coherence across the field, consistent with the central velocity spread seen in the P--V diagram (Fig.~\ref{fig:pv}). This suggests that the velocity field is globally continuous across the region, rather than being dominated by distinct, unrelated narrow components at each position.

The channel maps (Fig.~\ref{fig:cm}) further support this interpretation by showing a continuous spatial migration of emission with velocity, without evidence for sharply separated kinematic structures. These results are consistent with the \ion{H}{i} emission tracing a large-scale coherent envelope embedded in a structured ISM, rather than a simple superposition of unrelated discrete clouds.

\section{Dust properties}
\label{sec:photometry}

\begin{figure*}
\includegraphics[width=2.0\columnwidth]{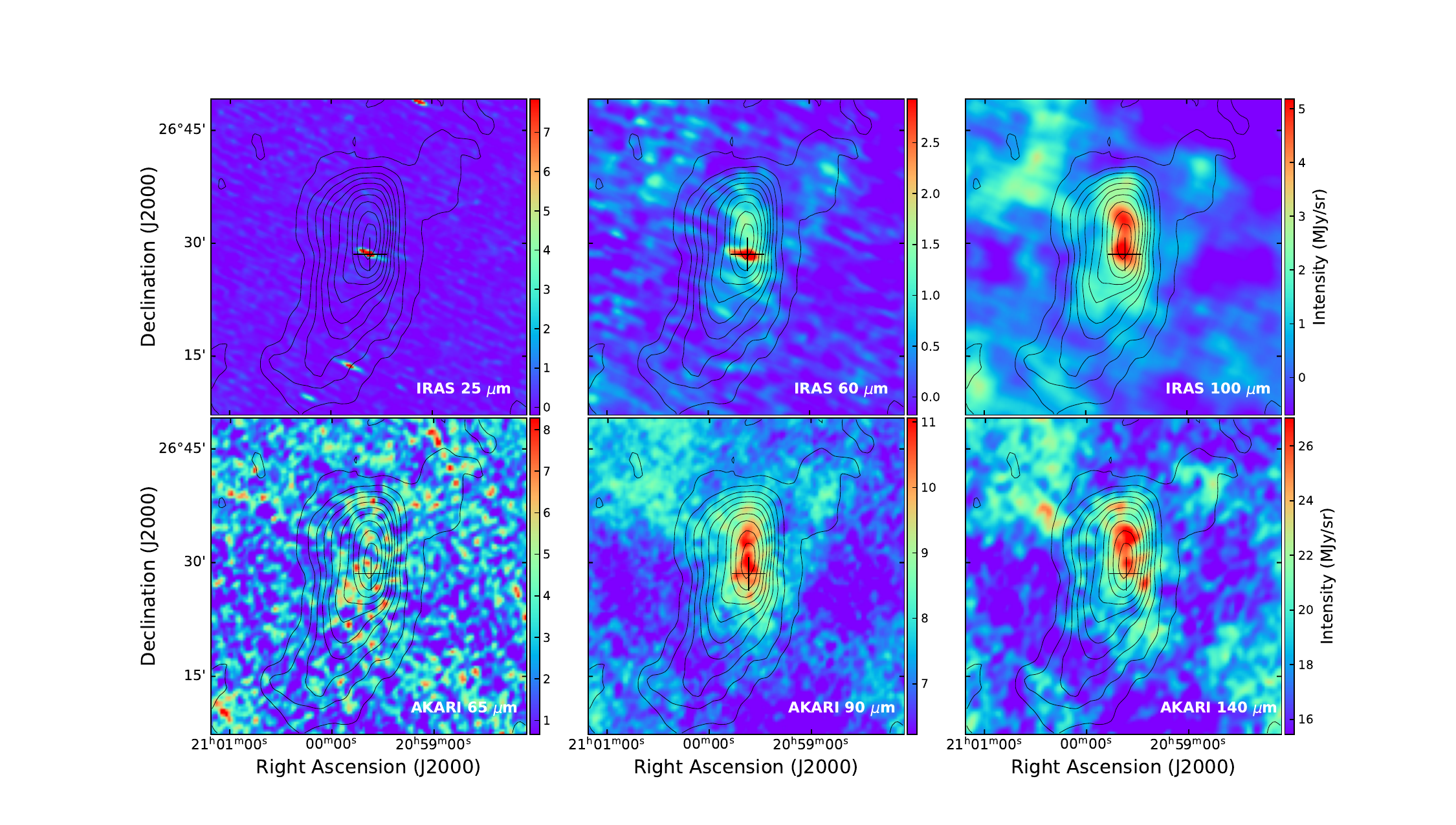}
\caption{
Multiwavelength infrared images of the circumstellar environment around V420\,Vul. In all panels, the background color-coded maps represent the infrared surface brightness in units of $\mathrm{MJy\,sr^{-1}}$. The overlaid black contours represent the local \ion{H}{i} integrated intensity map. The black cross in each panel marks the position of V420\,Vul. Note: the elongated shape of the $25\,\mu\mathrm{m}$ emission is a scanning artifact (see Appendix~\ref{sec:photometry}).
}
\label{fig:ir}
\end{figure*}

To perform a multiband comparison, we examined the IRAS and AKARI infrared data from 25 to 160\,$\mu\mathrm{m}$. We note that the 25\,$\mu\mathrm{m}$ map is affected by prominent linear scanning artifacts that artificially elongate the emission shape. Meanwhile, the 160\,$\mu\mathrm{m}$ map lacks discernible structures correlated with the \ion{H}{i} envelope and provides no additional constraints on the outflow morphology. Consequently, the 160\,$\mu\mathrm{m}$ band is excluded from the final compilation, while the 25\,$\mu\mathrm{m}$ image is included in Figure~\ref{fig:ir}.

Aperture photometry on the IRAS and AKARI images was performed using polygonal apertures tracing the \ion{H}{i} shell. The background-subtracted flux densities (in Jy) were obtained by integrating the surface brightness over the source mask after subtracting a median background measured in a clean adjacent region. Results are listed in Table~\ref{tab:dust_properties}.

\subsection{Dust temperature}

The characteristic dust temperature, $T_{\mathrm{d}}$, was derived from the flux ratio of two FIR bands ($60/100\,\mu\mathrm{m}$ for IRAS and $90/140\,\mu\mathrm{m}$ for AKARI). Assuming a modified blackbody emission modified by a power-law dust opacity ($\kappa_\nu \propto \nu^{\beta}$), the theoretical flux ratio can be expressed as

\begin{equation}
R_{\mathrm{th}} = \frac{F_{\nu_1}}{F_{\nu_2}} = \left(\frac{\nu_1}{\nu_2}\right)^{\beta} \frac{B_{\nu_1}(T_{\mathrm{d}})}{B_{\nu_2}(T_{\mathrm{d}})},
\end{equation}

where $B_{\nu}(T)$ represents the Planck function,

\begin{equation}
B_{\nu}(T) = \frac{2h\nu^{3}}{c^{2}} \frac{1}{\exp\left(\frac{h\nu}{k_{\mathrm{B}}T}\right)-1}.
\end{equation}

Here, $h$ is Planck's constant, $c$ is the speed of light, and $k_{\mathrm{B}}$ is Boltzmann's constant. By equating the observed ratio ($R_{\mathrm{obs}} = F_{\nu_1}/F_{\nu_2}$) to $R_{\mathrm{th}}(T_{\mathrm{d}})$, we solved for $T_{\mathrm{d}}$ numerically. Following the standard convention for AGB circumstellar dust grains, an opacity index of $\beta = 1.5$ was adopted throughout the calculations.

\subsection{Dust mass}

The total dust mass, $M_{\mathrm{d}}$, was inferred from the longer-wavelength flux density ($100\,\mu\mathrm{m}$ for IRAS and $140\,\mu\mathrm{m}$ for AKARI) according to the standard formulation:

\begin{equation}
M_{\mathrm{d}} = \frac{F_{\nu_2}\, D^{2}}{\kappa_{\nu_2}\, B_{\nu_2}(T_{\mathrm{d}})},
\end{equation}

where $D$ is the distance to V420\,Vul (as derived in Sect.~\ref{dis}). The dust mass absorption coefficient $\kappa_{\nu}$ at $100\,\mu\mathrm{m}$ was taken as $25.0~\mathrm{cm^{2}\,g^{-1}}$ following \citet{1983QJRAS..24..267H}. The corresponding value at $140\,\mu\mathrm{m}$ ($15.4~\mathrm{cm^{2}\,g^{-1}}$) was then derived assuming a standard FIR power-law dust opacity index of $\beta = 1.5$.

As summarized in Table~\ref{tab:dust_properties}, the AKARI photometry yields an observed flux ratio of $R_{\mathrm{obs}} = 0.4$, corresponding to a dust temperature of $T_{\mathrm{d}} = 19.4\,\mathrm{K}$ and a total dust mass of $M_{\mathrm{d}} \approx 0.8\,\mathrm{M}_{\sun}$. For comparison, the IRAS data yield a lower ratio of $0.3$, shifting the derived temperature upwards to $27.7\,\mathrm{K}$ and reducing the inferred dust mass to $\approx 0.1\,\mathrm{M}_{\sun}$. These discrepancies in both temperature and mass stem primarily from the differing spatial resolutions and imaging performance of the two instruments. While the coarser resolution of IRAS biases its measurements toward the warmer, compact components of the inner envelope, the superior spatial resolution and sensitivity of the AKARI FIR bands efficiently resolve and capture the diffuse, cold dust ($\sim 20$\,K) extended in the outer regions \citep{2007PASJ...59S.389K}. This naturally lowers the integrated temperature while recovering the bulk of the dust mass.

\begin{table}[t]
\caption{Far-infrared photometry and derived circumstellar dust properties for V420\,Vul.}
\label{tab:dust_properties}
\centering
\begin{tabular}{lcccc}
\hline\hline
\noalign{\smallskip}
Instrument & $\nu_1$\,/\,$\nu_2$ & $F_{\nu_1}$\,/\,$F_{\nu_2}$ & $T_{\mathrm{d}}$ & $M_{\mathrm{d}}$ \\
 & ($\mu\mathrm{m}$) & (Jy) & (K) & ($\mathrm{M}_{\sun}$) \\
\noalign{\smallskip}
\hline
\noalign{\smallskip}
IRAS  & $60$\,/\,$100$ & $9.5$\,/\,$30.6$ & $27.7$ & $0.1$ \\
AKARI & $90$\,/\,$140$ & $19.5$\,/\,$50.9$ & $19.4$ & $0.8$ \\
\noalign{\smallskip}
\hline
\end{tabular}
\tablefoot{The observed flux ratios ($F_{\nu_1}/F_{\nu_2}$) correspond to $0.3$ and $0.4$ for IRAS and AKARI, respectively. A dust opacity index of $\beta = 1.5$ is assumed for both datasets.}
\end{table}

\end{appendix}
\end{document}